\documentclass[10pt, final, journal, letterpaper,oneside, twocolumn]{IEEEtran}

\usepackage{amssymb}
\usepackage[cmex10]{amsmath}
\usepackage{amsfonts}
\usepackage{bbm}

\usepackage{amssymb}

\usepackage{pdfsync}

\usepackage{cite}

\ifCLASSINFOpdf
   \usepackage[pdftex]{graphicx}
   \graphicspath{{./Figures/}}
   \DeclareGraphicsExtensions{.pdf}
\else
\fi
\usepackage[cmex10]{amsmath}
\usepackage{amsfonts}

\begin{document}
%
\title{Femtocell Architectures with Spectrum Sharing for Cellular Radio Networks}
%
%
%


\author{\IEEEauthorblockN{Brett~Kaufman, \IEEEmembership{Student~Member,~IEEE,} Jorma~Lilleberg, \IEEEmembership{Senior~Member,~IEEE,}
\\and~Behnaam~Aazhang, \IEEEmembership{Fellow,~IEEE}}

\thanks{B. Kaufman, J. Lilleberg, and B. Aazhang are jointly with the Center for Multimedia Communication at Rice University and the Centre for Wireless Communication at the University of Oulu, Finland.  J. Lilleberg is also with Renesas in Finland.  This work is funded in part by NSF, the Academy of Finland through the Co-Op grant, and by Renesas through a research contract.}}
\maketitle

\vspace{-55pt}
\begin{abstract}
Femtocells are an emerging technology aimed at providing gains to both network operators and end-users.  These gains come at a cost of increased interference, specifically the cross network interference between the macrocell and femtocell networks.  This interference is one of the main performance limiting factors in allowing an underlaid femtocell network to share the spectrum with the cellular network.  To manage this interference, we first propose a femtocell architecture that orthogonally partitions the network bandwidth between the macrocell and femtocell networks.  This scheme eliminates the cross network interference thus giving the femtocells more freedom over their use of the spectrum.  Specifically, no interference constraint is imposed by the cellular network allowing femto users to transmit at a constant power on randomly selected channels. Although simple, this scheme is enough to give gains up to 200\% in sum rate.  

We then propose a second architecture where both networks share the bandwidth simultaneously.  A femtocell power control scheme that relies on minimal coordination with the macrocell base station is used in conjunction with an interference sensing channel assignment mechanism. These two schemes together yield sum rate gains up to 200\%.   We then develop a technique for macro users to join a nearby femtocell and share a common channel with a femtocell user through the use of successive interference cancellation.  By adding this mechanism to the power control and channel assignment schemes, we show sum rate gains over 300\% and up to 90\% power savings for macrocell users.  

\end{abstract}



%
\IEEEpeerreviewmaketitle

\vspace{-15pt}
\section{Introduction}
\label{sec:intro}
Consumer markets are becoming saturated with personal wireless devices like tablet computers, smartphones, and gaming systems.  The affordability of these devices results in an average user owning multiple devices, each with their own demand for wireless service.  This high degree of connectivity coupled with the constant demand for higher data rates is putting an enormous strain on todays wireless networks.  Recent auctions of the television broadcast spectrum \cite{FCC_AUCTION} show the high cost and difficulty in acquiring new spectrum.  More practical solutions will involve novel techniques in the way wireless devices access the network and share the wireless resources.

One emerging solution is consumer installed femtocells in indoor environments \cite{Claussen2008An-Overview-of-}.  Femtocells have been receiving considerable attention recently in both academic works \cite{Femtocell_Survey} as well as cellular standards like 3GPP \cite{3GPPstan}.  From the network operatorÕs perspective, both spectral efficiency and user capacity can increase.  End-users connecting to femtocells can experience improved signal quality and significant power savings.  The biggest obstacle in achieving these gains is the interference management between the femtocell users and the traditional cellular users \cite{femtoforum}.  Two survey works were done focusing on interference avoidance techniques in OFDM systems \cite{lopez_interference} and cellular systems \cite{andrews_int_can}.  Similar to cellular networks, power control in the femtocell network has become a commonly used approach \cite{andrews_powercontrol,Han-Shin-Jo2009Interference-Mi}.  Other techniques like dynamic channel assignment \cite{hybrid_freqassign} and novel frequency reuse techniques have also emerged \cite{FFR_infocom}.  One promising area of focus is in interference cancellation \cite{int_can_ahls,andrews_int_can}.  Specifically, successive interference cancellation was considered for cellular networks with femtocells in a recent survey work \cite{int_can_4G}.  

In this work we develop two femtocell architectures.  The first one allocates orthogonal partitions of the network bandwidth to the macrocell and femtocell networks.  In doing so, femtocells are able to operate within their allocated spectrum without any regard to the cellular base station or its macrocell users, however inter-femtocell interference becomes the dominant limiting factor of the femtocell network.  We show in our work that a scheme that uses constant transmit power with random channel selection is sufficient to provide sum rate gains up to 200\%.  

In the second part of our work, we propose a femtocell architecture where the entire spectrum is shared simultaneously with both networks.  Due to the shared spectrum, the existing cellular network and specifically the base station, impose a constraint on the total allowed interference from femtocell users accessing the spectrum.  We develop a power control scheme for femtocell users that relies on minimal coordination with the cellular base station.  We then use an interference sensing channel assignment technique to establish links between femtocell users and the femtocell access point.  We show that these two techniques combined can yield sum rate gains up to 200\%.

To further improve the interference management in our proposed shared spectrum architecture, we develop a decision rule for conditions in which macrocell users should connect to a nearby femtocell instead of the cellular base station.  We use successive interference cancellation to allow a macrocell user and femtocell user to share a common channel and establish a link with the femtocell access point.  Successive interference cancellation has been shown as a feasible technique in OFDM networks for both uncoded \cite{SIC_OFDM} and coded systems \cite{andrews_int_can}.  The performance of the successive interference cancellation depends largely on the channel estimation of the interfering signal so that it can be successfully subtracted from the desired signal of interest. We use perfect cancellation to upper bound the performance of the system and then show that the network suffers only minor performance loss for cancellation errors up to 12\%.  

Combining the femtocell user power control and channel assignment with this scheme that allows macrocell users to join nearby femtocells can increase the sum rate gains to over 300\%.  Additionally, femtocell users and macrocell users can have up to 70\% and 90\% power savings.

The remainder of the paper is organized as follows: Section \ref{sec:netmodel} defines the two-tier macrocell-femtocell model under consideration.  In Section \ref{sec:splitspec} we present a split spectrum femtocell architecture where the network's bandwidth is partitioned orthogonally between macro and femto users.  Section \ref{sec:sharespec} then presents a shared spectrum femtocell architecture where successive interference cancellation is used to allow macro users to join a nearby femtocell.  Concluding remarks on the work will be presented last.  

\section{Network Model}
\label{sec:netmodel}
\begin{figure}
  \includegraphics[width=0.48\textwidth]{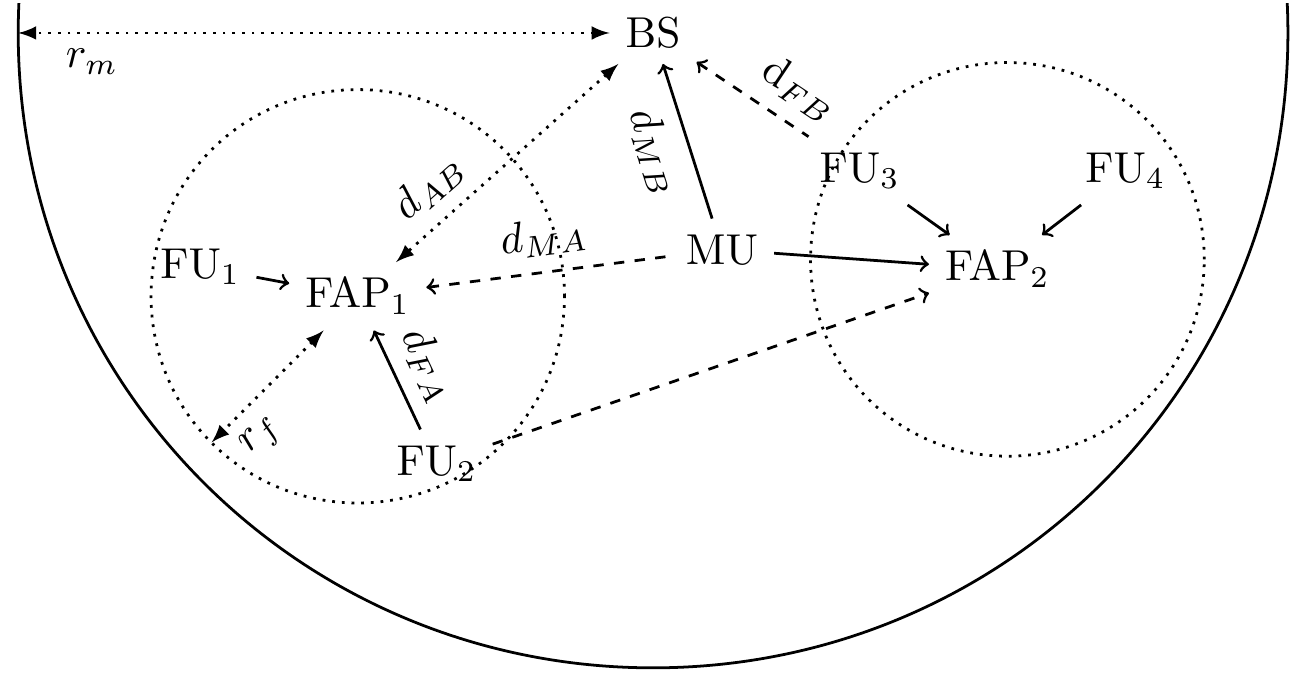}
\caption{Example topology showing the various users and random distances in the model.}
\label{fig:sysmodel}
\end{figure}

We refer to Fig.~\ref{fig:sysmodel} as we describe our network model.  We separate the details of our model in terms of the infrastructure, users, frequency resources, and channel model.  Our work analyzes the performance of the network in the uplink frame and thus the details will be given with respect to that mode.  
\subsection{Infrastructure Model}
We consider a single circular macrocell of radius $r_m$ with a base station (BS) located at the center.  We assume that the BS has a single omnidirectional antenna and serves users located within the cell boundary.  We refer to the region enclosed by the cell as $A$ and denote the size of the coverage area of the BS as $|A|$.

We consider the macrocell to be underlaid with circular femtocells of radius $r_f$.  There is a femtocell access point (FAP) located at the center of each femtocell with a wired backhaul connection.  We assume the FAPs are uniformly distributed inside $A$ according to a two-dimensional spatial poisson point process $\Omega_{f}$ with intensity $\lambda_{f}$.  For practical considerations, we will use the average number of femtocells per macrocell, found as $N_{f} = \lambda_{f} |A|$.  

We note that the random uniform location of femtocells makes it equally likely that two femtocells are located on opposite sides of the cell or at the same location.  Our work considers all femtocell locations as feasible and makes no assumptions about location planning or optimization by the service provider.   Finally, we do not assume that femtocells are networked together and thus have no knowledge of each others activity or location.  

We assume that the BS and FAP can access the same core network and thus both are deployed by the same service provider, or use common hardware.  The network is considered to be in uplink mode thus both the BS and FAP will be receiving data from their respective users.  Common access to the core network can be used to synchronize both networks to the uplink frame and exchange relevant control signaling.  We note that the common access to the core network does not imply that the femtocells are centrally controlled by the cellular base station nor does the BS and FAP work together for joint decoding as proposed in related works.  

\subsection{User Model}
\label{subsec:users}
We assume that there are $M$ macro users (MU) uniformly distributed inside the macrocell.  The macro user's location is independent of the femtocell network and thus it is feasible that one or more macro users could be located within a single femtocell.  A minimum SINR of $\beta_{M}$ is required for a macro user to establish a link with the BS.  Inside every femtocell, we assume $F$ uniformly distributed femto users (FU).  A minimum SINR of $\beta_{F}$ is required for a femto user to establish a link with the FAP.  We assume both types of users have a single antenna and similar hardware enabling them to access either the BS or FAP.  With the network in the uplink mode, both macro and femto users will be transmitting and thus will not receive interference from each other.  Finally, both types of users will utilize standard cellular control signaling to establish a link with their respective access point.  

\subsection{Frequency Resources}
\label{subsec:chans}
We consider the network's uplink bandwidth to be divided into a set $\mathcal{C} = \{C_{1},\dots, C_{N_{C}} \}$ of $N_{C}$ orthogonal channels and refer to the $n$'th given channel as $C_{n}$.  In the context of this work, a channel could be either a frequency or time block as in OFDM or TDMA systems.  We assume that each of the $N_{C}$ channels is allocated to just one macro user, which can be expressed by $N_{C} = M$.  This assumption is in place to ensure that all frequency resources are actively in use and there are no free resources to be exchanged between the two networks.  We make a similar assumption for femto users such that we assume a given channel $C_{n}$ can only be allocated once per femtocell.  In our work, we will consider two spectrum sharing schemes:

\paragraph{Split Spectrum Scheme}In this scheme, we assume that the $N_{C}$ orthogonal channels will be partitioned between the macrocell and femtocell networks.  We use the parameter $\gamma$ to denote the number of channels allocated to the femtocell network.  As a minimum level of performance, we assume that the number of channels allocated to the femtocell network is at least equal to the number of femto users in each femtocell, that is we require $\gamma \geq F$.  Finally, we assume that all femto users in this scheme transmit at a constant power level.  There are no interference constraints imposed on the femtocell network by the macrocell network, and femto users are located near the FAP, thus a constant transmit power is feasible.  

\paragraph{Shared Spectrum Scheme}In this scheme, both macro and femto users will use the same channels simultaneously.  By sharing the spectrum, the macrocell base station will receive interference from femto users and similarly, the femtocell access point will receive interference from macro users.  Due to fluctuations in interference and noise effects in the network, infrastructure in today's wireless networks are designed to operate within a minimally varying interference temperature \cite{Martin-Sacristan2009On-the-Way-Towa}.  To model the amount of allowed interference at the base station, we assume that there is a margin $\kappa_{M}$ in the SINR at the BS.  This margin give the amount of allowed SINR variation at the access point due to interference before a link can no longer be maintained.  

\subsection{Channel Model}
We consider three arbitrary users: a transmitter $i$, a receiver $j$, and an interferer $k$.  We assume a pathloss dominated channel with additive white Gaussian noise.  The pathloss is determined by the Euclidian distance $d_{ij}$ between two users $i$ and $j$ and the pathloss exponent $\alpha$.  We are primarily interested in the power of user's signals and the corresponding SINR of their links and thus define user $j$'s SINR on a given channel $C_{n}$ as
\begin{equation}
\Gamma_{j,n} = \dfrac{P_{T_{i}}d_{ij}^{-\alpha}}{\sum\limits_k P_{T_{k}}d_{kj}^{-\alpha} + \sigma^2}
\label{eq:chanmodel}
\end{equation}
where $P_{T_{i}}$ is the power used by the transmitter and $d_{ij}^{-\alpha}$ is the pathloss for the link between the transmitter and receiver.  Similarly, $P_{T_{k}}$ is the power used by the $k$'th interferer and $d_{kj}^{-\alpha}$ is the pathloss between the $k$'th interferer and the receiver.  We assume that all users observe the same noise power of $\sigma^2$.  We will use the subscripts $M$, $F$, $B$, and $A$ to denote the different parameters for the macro user, femto user, macrocell base station, and femtocell access point.  Using our distance based pathloss model, modeling the interference on the various links is equivalent to varying the pathloss exponent.  For femto users communicating with the femtocell access points, we use $\alpha$ as the pathloss exponent.  We use an exponent of $\psi$ for macro users when they interfere with the femtocell access points or when they connect to a nearby femtocell access point.  Finally, we use $\phi$ as the exponent for macro user links with the base station as well as the femto user interference to the base station and other femtocell access points.

\section{Split Spectrum Femtocell Architecture}
\label{sec:splitspec}
In this section, we develop the femtocell architecture when the two networks are allocated disjoint partitions of the spectrum.  By splitting the spectrum, there will be no cross network interference.  The macrocell network will perform as if the femtocells were not even there.  However, the femtocell network performance will be completely determined by the interference effects from neighboring femtocells.  All femto users use the same constant transmit power, so the only task required at each femtocell access point is to assign channels to its $F$ femto users.  

Recall we assume that $\gamma \geq F$.  When $\gamma = F$, each femtocell will use the same $\gamma$ channels and each FAP will receive interference from every other femtocell.  When $\gamma > F$, each FAP will select $F$ channels with uniform probability of $1/\gamma$.  By randomly selecting channels like this, the frequency resources of adjacent femtocells will not always be identical.  As $\gamma$ increases, the probability that two neighboring femtocells use the same channel decreases.  This in turn reduces the interference that the FAP sees on each channel.  
\begin{table}
\center
\caption{Network Simulation Parameters}
\label{tab:params}
\begin{tabular}{ll}
\hline\noalign{\smallskip}
System Paramter & Value \\
\noalign{\smallskip}\hline\noalign{\smallskip}
Average number of femtocells ($N_{f}$) & [0,40] \\
Number of channels ($N_{C}$) & 25 \\
Number of macro users ($M$) & 25 \\
Number of femto users per femtocell ($F$) & 5 \\
Minimum macro user SINR ($\beta_{M}$) & 20 dB \\
Minimum femto user SINR ($\beta_{F}$) & 25 dB \\
Noise Power ($\sigma^{2}$) & -95 dBm \\
Macrocell radius ($r_{m}$) & 400 m \\
Femtocell radius ($r_{f}$) & 30 m \\
Pathloss exponents ($\alpha$, $\psi$, $\phi$)& 2, 3, 3.5 \\ 
\noalign{\smallskip}\hline
\end{tabular}
\end{table}

\begin{figure}
  \includegraphics[width=0.48\textwidth]{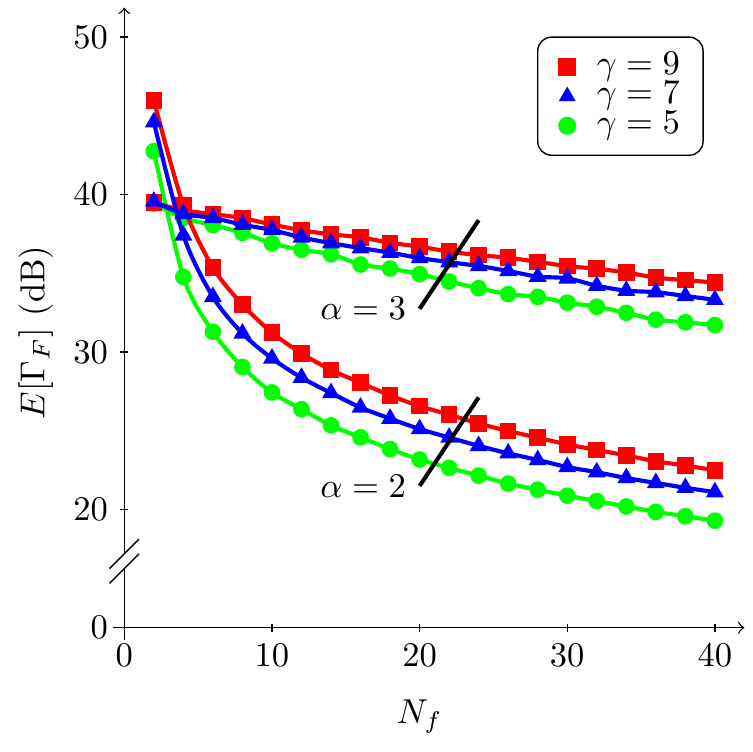}
\caption{The average femto user SINR ($E[\Gamma_{F}]$) versus the average number of femtocells ($N_{f}$) per macrocell.  }
\label{fig:avgsinr}
\end{figure}

To see this effect, we simulated the network shown in Fig.~\ref{fig:sysmodel} in Matlab for 5000 random topologies.  The network parameters used for evaluating the network are shown in Table \ref{tab:params}.  We can see the effect of $\gamma$ on a femto user's performance in Fig.~\ref{fig:avgsinr}
where we plot the average SINR for a given femto user, $E[\Gamma_{F}]$, versus the average number of femtocells per macrocell.  We can see that increasing $\gamma$ has only a minimal effect on a femto user's SINR.  

To quantify the combined performance of the two networks, we calculate the network sum rate as
\begin{equation}
\mathcal{R}_{sum} = \mathcal{R}_M + \mathcal{R}_F,
\label{eq:sumrate}
\end{equation}
where $\mathcal{R}_M$ is the sum rate of the macrocell network and $\mathcal{R}_F$ is the sum rate of the femtocell network.  Recall we assume $N_C = M$ so that when there are no femtocells in the network, all of the channels will be actively in use by macro users.  Furthermore, as $\gamma$ channels are allocated to the femtocell network, only $N_{C}-\gamma$ macro users can be served by the BS with an SINR of $\beta_{M}$.  The zero rate of the $\gamma$ unserved macro users will decrease the sum rate of the macrocell network.  Based on the above, we can write
\begin{equation}
\mathcal{R}_M = (N_{C}-\gamma) \log_{2}\left(1+ \beta_{M} \right),
\label{eq:mrate}
\end{equation}
which is the sum rate of the macrocell network.  

We can calculate the femtocell component of the mean sum rate in a similar manner.   There are $N_f$ femtocells per macrocell with $F$ femto users per each femtocell.  We know that users are assigned a given channel $C_{n}$ with probability $1/\gamma$.  Thus for a given network realization, there will be $FN_{f}/\gamma$ femto users per channel $C_{n}$ in which user $j$ has an instantaneous link quality of $\Gamma_{j,n}$.  By summing the rate of all users over all $\gamma$ channels, we can write 
\begin{equation}
\mathcal{R}_F =  \sum_{n =1}^{\gamma}\sum_{j=1}^{^{FN_{f}}/_{\gamma}} \mathbbm{1}_{\beta_{F}}(\Gamma_{j,n})\log_{2}\left(1+ \beta_{F}\right),
\label{eq:frate}
\end{equation}
which is the sum rate of the femtocell network.  We note the use of the indicator function where $\mathbbm{1}_{\beta_{F}}(\Gamma_{j,n})  =1$ if $\Gamma_{j,n} \geq \beta_{F}$ and $0$ otherwise.  Using (\ref{eq:mrate}) and (\ref{eq:frate}), we are able to calculate the mean achievable sum rate in (\ref{eq:sumrate}). 

In order to quantify the combined performance of the macro and femtocell networks, we calculate the gain in the network sum rate.  We measure the gain with respect to a macro user only network where $\gamma = 0$ and there are $N_{C} = M$ active macro users each with a SINR of $\beta_{M}$.  Thus using (\ref{eq:sumrate}), we can write
\begin{equation}
\mathcal{R}_{gain}^{split} = \dfrac{\mathcal{R}_{sum} - M \log_{2}(1+ \beta_{M})}{M \log_{2}(1+ \beta_{M})},
\label{eq:sumrategainsplit}
\end{equation}
which is the gain in the network sum rate for the split spectrum scheme.  As $\gamma$ increases, the number of macro users who are no longer served by the base station also increases, thus lowering the macrocell component of the sum rate.  However, a given channel that was allocated to a single macro user can now be shared among multiple femto users in different femtocells.  This can potentially increase the femtocell component of the sum rate significantly.  

\begin{figure}
  \includegraphics[width=0.48\textwidth]{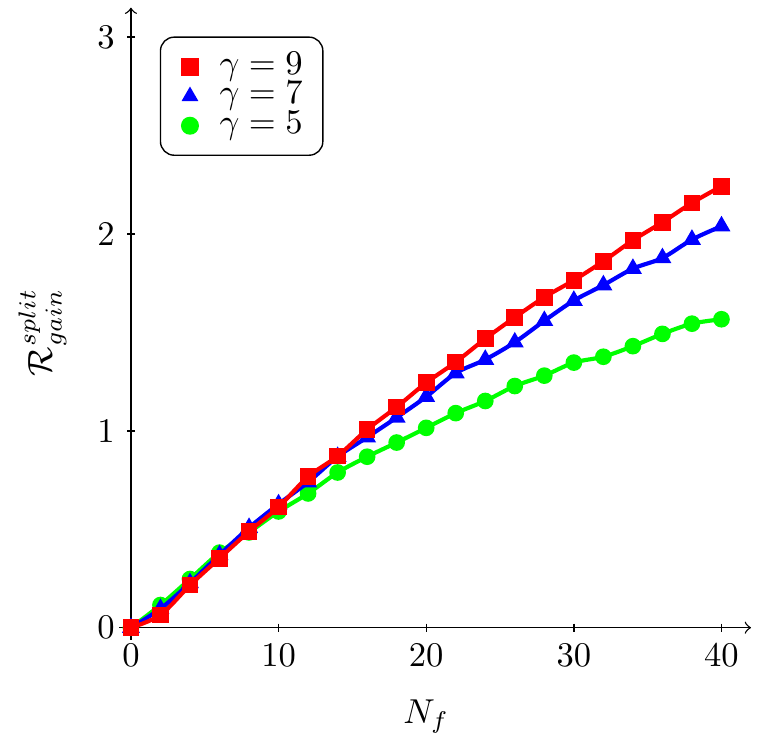}
\caption{The average network sum rate gain ($\mathcal{R}_{gain}^{split}$) versus the average number of femtocells ($N_{f}$).  }
\label{fig:tputgainsplit}
\end{figure}

In Fig.~\ref{fig:tputgainsplit}, we plot the average sum rate gain from (\ref{eq:sumrategainsplit}) versus the average number of femtocells per macrocell.  
We can immediately see that for all values of $\gamma$, there is a sum rate gain.  As discussed above, the reuse of a given channel by many femto users can provide significant benefits to the performance of the network as a whole.  The lowest curve corresponds to the scenario of $\gamma = F$ where every femtocell uses the same $\gamma$ channels for its $F$ femto users.  Thus every femtocell access point is receiving interference on every channel.  Despite that, the isolation of the femtocells from each other still enables gains to be achieved.  As $\gamma$ increases, the gains also increase.  However, $\gamma$'s affect on the gain for each additional channel allocated to the femtocell network decreases.  Looking at $\gamma = 5$, we see the curve approaching a maximum for large $N_{f}$.  At some point, the network will become saturated with femtocells and additional users cannot be supported.  The same trend will be observed for higher values of $\gamma$ at corresponding large values of $N_{f}$.

\section{Shared Spectrum Femtocell Architecture}
\label{sec:sharespec}
We now develop the architecture for a more interesting, and more complicated, network where macro and femto users share the same spectrum.  
We will specify the details for the power control and channel assignment.  Additionally, we will provide a scheme where successive interference cancellation can be used at the femtocell access points to enable macro users to join nearby femtocells.    
\subsection{User Power Control}
Power control needs to be performed for both macro and femto users.  Recall that there exists a margin $\kappa_M$ in the SINR at the base station to account for some allowed level of interference.  We can see the effects of the macro user power control by looking at the SNR of a given macro user uplink with the BS, where after rearranging terms, gives a bound on the transmit power of macro users as
\begin{eqnarray}
\dfrac{P_{T_{M}} d_{MB}^{-\phi}}{\sigma^2} &\geq& \kappa_M \beta_M \nonumber
\\
P_{T_{M}} &\geq& \kappa_M  \beta_{M} \sigma^2 d_{MB}^{\phi},
\label{eq:ptm_bound}
\end{eqnarray}
where $d_{MB}^{-\phi}$ is the pathloss between a macro user and the BS and $P_{T_M}$ is the transmit power of the macro user.  If we assume that macro users transmit at the required minimum just found, then by looking at the SINR of a given channel being shared with femto users, and after rearranging terms
\begin{eqnarray}
\dfrac{P_{T_{M}} d_{MB}^{-\phi}}{\sum\limits_k P_{T_{k}}d_{kB}^{-\phi} + \sigma^2} &\geq& \beta_M \nonumber
\\
\sigma^2(\kappa_M -1) &\geq& \sum\limits_k P_{T_{k}}d_{kB}^{-\phi},
\label{eq:int_bound}
\end{eqnarray}
gives an upper bound on the total allowed interference power at the base station in terms of the interference margin $\kappa_M$.  We note that $P_{T_{k}}$ and $d_{kB}^{-\phi}$ are the values for the transmit power and pathloss from the $k$'th interfering co-channel femto user.  Recall from the assumptions in Section \ref{subsec:chans} that each femtocell access point can only allocate a given channel once among its $F$ users.  Thus each value of $k$ in (\ref{eq:int_bound}) represents a femto user from different femtocells.  If we divide $\sigma^2(\kappa_M -1)$ by the average number of femtocells $N_f$, we can calculate the amount of interference power allowed from a given femto user per femtocell.  Using our pathloss channel model, we know that the interference from a femto user to the BS is simply $P_{T_F}d_{FB}^{-\phi}$.  Combining these concepts, we can write
\begin{equation}
\dfrac{\sigma^2(\kappa_M -1) d_{FB}^{\phi}}{N_f} \geq P_{T_F},
\label{eq:FU_pt_bound}
\end{equation}
which is an upper bound for the transmit power of a typical femto user on a given channel.  

We assume that each femtocell's FAP manages the power control for its own users but the process is aided by the BS.  The amount of overhead necessary for the BS to learn $d_{FB}$ for every femto user in each femtocell could be quite high.  With an aim to keep the overhead to a minimum, a close approximation of $d_{FB}$ can be made.  We assume a worst case location for a typical femto user as being at the point on the edge of a femtocell closest to the BS.  We illustrate this in Fig.~\ref{fig:sysmodel}, where we show a femto user FU$_3$ on the edge of FAP$_2$.  Due to the relative small size of the femtocell, we can approximate each femto users distance by $d_{AB} - r_f$, the difference of the distance from the FAP to the BS and the femtocell radius.  Because the FAPs are stationary, the overhead for the BS to know the distance to the FAPs is low.  Thus for any femto user in a given femtocell, its own distance to the BS will always satisfy $d_{FB} \geq d_{AB} - r_f$.  Combining this with (\ref{eq:FU_pt_bound}), we get
\begin{equation}
\dfrac{\sigma^2(\kappa_M -1) (d_{AB}-r_f)^{\phi}}{N_f} \geq P_{T_F},
\label{eq:FU_pt_bound2}
\end{equation}
which gives an upper bound for the transmit power for all femto users served by a common FAP.  We assume that the BS knows $d_{AB}$ for each femtocell and it knows $N_F$, and thus can set a maximum transmit power level for each femtocell.  Using (\ref{eq:FU_pt_bound2}) as a maximum power constraint, we assume FAPs employ standard power control techniques with their femto users.  

\subsection{Macrocell to Femtocell Handover}
We showed in the previous section how $\kappa_M$ constrains the total allowed femtocell interference on each channel at the BS.  By doing so, macro users will always be able to maintain their required SINR threshold with the BS and have no reason in terms of link reliability to connect to a nearby femtocell. In order to motivate a handover procedure in which macro users can uplink to a nearby femtocell access point rather than the BS, we consider potential power savings.  Depending on the topology of the network, a macro user could potentially use less power to join a nearby femtocell.  

We can define a simple decision rule in which a macro user should join a nearby femtocell if the transmit power needed to uplink to the FAP, $P_{T_M}^{*}$, is less than the power needed to transmit to the BS, $P_{T_M}$.  If we write the required SINR constraint for a macro user uplinking to a FAP on a given channel and rearrange terms, 
\begin{eqnarray}
\dfrac{P_{T_{M}}^{*} d_{MA}^{-\psi}}{\sum\limits_k P_{T_{k}}d_{kA}^{-\phi} + \sigma^2} \geq \beta_M \nonumber
\\
P_{T_{M}}^{*} \geq  \beta_{M} \Big(\sum\limits_k P_{T_{k}}d_{kA}^{-\phi} + \sigma^2\Big)d_{MA}^{\psi},
\label{eq:FA_pt_bound}
\end{eqnarray}
we get a bound for the transmit power needed to reach the FAP.  This bound is proportional to the pathloss of the channel from the macro user to the FAP, as well as the received interference at the FAP.  We note that the interference contribution here comes from co-channel femto users located in other femtocells.  Thus if we define the decision rule for which a macro user should connect to a FAP as $P_{T_M} > P_{T_M}^{*}$ and use the minimum powers derived in (\ref{eq:ptm_bound}) and (\ref{eq:FA_pt_bound}), after rearranging terms we get
\begin{equation}
d_{MB}^{\phi} > \dfrac{\Big(\sum\limits_k P_{T_{k}}d_{kA}^{-\phi} + \sigma^2\Big) d_{MA}^{\psi}}{\kappa_M \sigma^2}
\label{eq:decision_rule}
\end{equation}
which gives the decision rule in terms of the network parameter $\kappa_M$, the co-channel interference at the FAP, and the pathloss of the two different links.  We assume that there is a mechanism in place in which macro suers can learn the pathloss of those two links \cite{bkaufman_j1}. 

Using the decision rule in (\ref{eq:decision_rule}), macro users can utilize standard handover techniques with a nearby FAP to join its femtocell.  Once within the femtocell, the macro user can be power controlled like a femto user and utilize the wired backhaul link to relay its data.  Recall from Section \ref{subsec:users} that each FAP can only support $F$ links, one link for each of the $F$ femto users in the femtocell.  As a solution, we intend for a macro user who is admitted to a nearby femtocell to share a channel simultaneously with a femto user in a multiple access method.  We propose successive interference cancellation as the method in which this shared channel can be sustained.   

\subsection{Successive Interference Cancellation}
We utilize successive interference cancellation (SIC) at the femtocell access point to allow a femto user and a macro user to share a common channel.  SIC works by canceling the most recently decoded user's signal from the remaining signal by re-encoding the decoded message, modulating the new signal based on channel estimates, and subtracting it from the remaining signal.  As mentioned in the beginning of this work, the performance of SIC relies heavily on the channel estimation of the interfering signal.  Specifically, errors present in the amplitude or phase estimation used in the modulation step could result in an erroneous estimate of the most recently decoded signal.  

In this work, we assume the macro user is the primary user and the femto user is the interfering user.  The femto user is located in close proximity to the FAP and is often slow moving or stationary.  Thus the channel estimation for a femto user link should have a high probability of low error.  We follow the methodology in \cite{cover_thomas} to set up the two-user multiple access channel.  To begin, both the femto user and macro user transmit simultaneously.  By treating the macro user's signal as noise, the femtocell access point can decode the femto user first as long as the required SINR constraint
\begin{equation}
\dfrac{P_{T_{F}} d_{FA}^{-\alpha}}{\mathcal{I}_M + \sum\limits_k P_{T_{k}}d_{kA}^{-\phi}  + \sigma^2} \geq \beta_F,
\label{eq:sic1}
\end{equation}
is met.  We note that $\mathcal{I}_M = P_{T_{M}}^{*} d_{MA}^{-\psi}$ and use the notation $\mathcal{I}_M$ to clearly distinguish when the macro user's signal is considered as interference.  In a similar fashion, we define $\mathcal{I}_F = P_{T_{F}} d_{FA}^{-\alpha}$ to represent the femto users signal when it is interfering with the macro users signal.  As mentioned above, after decoding the first user's signal, it can be subtracted from the remaining signal.  We can see this process by looking at the required SINR constraint
\begin{equation}
\dfrac{P_{T_{M}}^{*} d_{MA}^{-\psi}}{\mathcal{I}_F - \widehat{\mathcal{I}_F} +  \sum\limits_k P_{T_{k}}d_{kA}^{-\phi} + \sigma^2} \geq \beta_M,
\label{eq:sic2}
\end{equation}
for the macro user, where $\widehat{\mathcal{I}_F}$ denotes the estimate of the now interfering femto user.  If the estimate of the interfering signal is free of error, that is $\mathcal{I}_F = \widehat{\mathcal{I}_F}$, then the macro user can meet its required SINR threshold.  However, if an error is present in the estimate, there is some nonzero probability that a macro user cannot be decoded successfully.  To consider this, we use the estimate $\widehat{\mathcal{I}_F} = \mathcal{I}_F - \mathcal{I}_\varepsilon$, where $\mathcal{I}_\varepsilon$ is the residual interference power left after an imperfect cancellation.  The amount of $\mathcal{I}_\varepsilon$ left after cancellation will have a direct impact on the SINR of the macro users signal.  In order to quantify this effect, we define $\mathcal{I}_\varepsilon$ as the amount of power necessary to cause an $\varepsilon$ decrease in the SINR of the macro users signal at the femtocell access point.  We can formally express this as 
\begin{equation}
\dfrac{P_{T_{M}}^{*} d_{MA}^{-\psi}}{\mathcal{I}_\varepsilon +  \sum\limits_k P_{T_{k}}d_{kA}^{-\phi} + \sigma^2} \triangleq (1-\varepsilon)\Gamma_M,
\label{eq:sic3}
\end{equation}
where $\Gamma_M$ is the measured SINR of the macro user at the FAP.  Thus using the steps shown above, SIC can be used to form a multiple access channel to allow a macro user and femto user to share a common channel after a handover has been made.  

\subsection{Channel Assignment}
At this stage, macro users have been either power controlled by the BS or admitted to a femtocell on a shared channel with a femto user.  Femto users have a maximum power imposed on them by the access point in their femtocell as found in (\ref{eq:FU_pt_bound2}). The final task that remains is for a FAP to assign channels to its femto users not already sharing a channel with a macro user.  It is in the best interest of the femto users in terms of power consumption and link outage to use the channels with the least amount of interference.  To accomplish this, we assume that each FAP measures the interference power on the channels that haven't already been allocated to a macro-femto user pair.  For convenience, we define the interference power of a given channel $C_n$ to be $\mathcal{I}_n$.  We know that of all $N_C$ total channels, at most $F$ of them can be shared by macro-femto user pairs after the handover process, leaving $N_{C} - F$ remaining channels for allocation.  Thus after measuring the interference powers, we assume each FAP maintains an ordered set of channels $\{C_1, C_2, ..., C_{N_C-F} \}$ such that $\mathcal{I}_1 < \mathcal{I}_2 < ... < \mathcal{I}_{N_C-F}$.  Given this ordered set, each FAP can then assign the least interfered channels to its femto users.  

\subsection{Results}
In this section we provide simulation results for the performance of the femtocell architecture described above.  We refer to Table~\ref{tab:params} as values for the various network parameters.  We will compare two different strategies as we present our results.  The first scheme we consider is one that uses the power control and channel assignment methods described above but does not allow a macro user to handover to a nearby femtocell.  We will refer to this as the PC scheme.  The second scheme we consider uses the same power control and channel assignment as the PC scheme but also uses the macro user handover process in which a macro-femto user pair can share a channel by use of successive interference cancellation.  We label this scheme as SIC for convenience.  

\begin{figure}
  \includegraphics[width=0.48\textwidth]{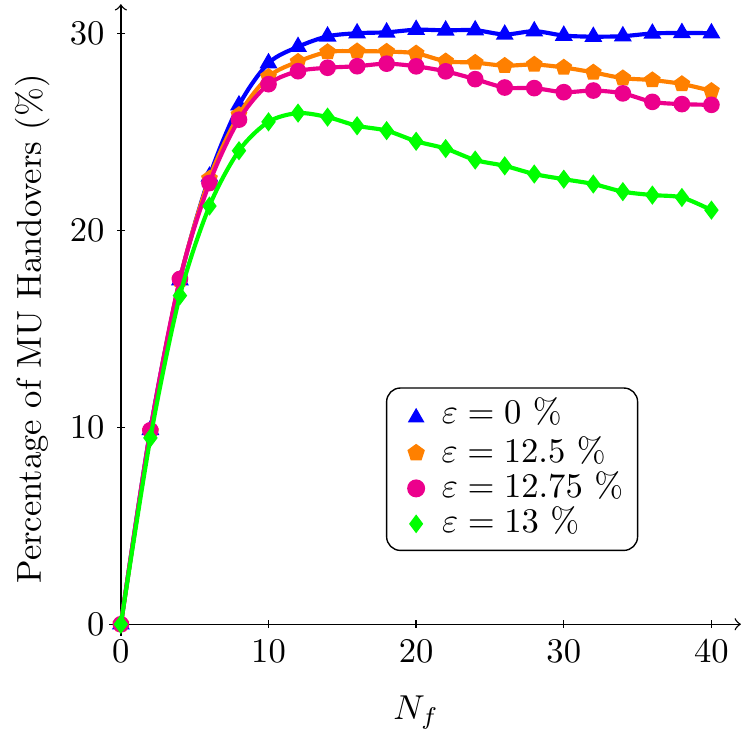}
\caption{The average number of macro user to femtocell handovers that occur in the network as a function of the cancellation error $\varepsilon$.}
\label{fig:handover_noerror}
\end{figure}

\begin{figure}
  \includegraphics[width=0.48\textwidth]{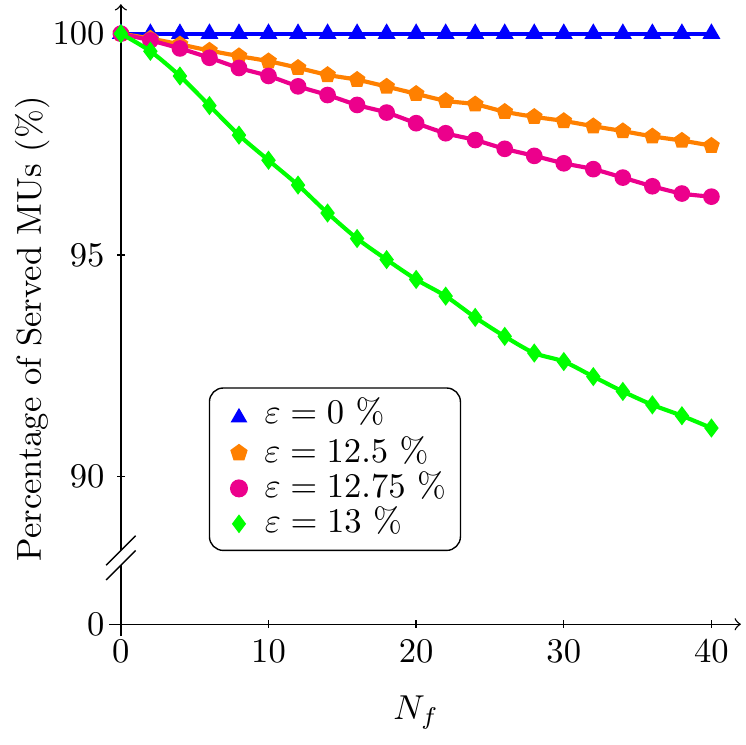}
\caption{The percentage of macro users who maintain their required SINR with the femtocell access point after performing a handover to join the femtocell.}
\label{fig:goodlinksMU}
\end{figure}

We begin our discussion by evaluating the performance of the handover process for a MU to join a nearby femtocell.  In Fig.~\ref{fig:handover_noerror}, the average number of successful macro user handovers versus the average number of femtocells is plotted.  
We consider a handover to be successful if a macro user can maintain their required SINR threshold $\beta_{M}$ after connecting to the femtocell.  Recall that $\varepsilon$ quantifies how much error occurs in the cancellation of the femto user's interfering signal from the macro user's signal.  For the case of $\varepsilon = 0$, the femto user's interfering signal is cancelled perfectly and thus no macro user who performs a handover will be in outage.  From the figure, we can see that with perfect cancellation a maximum handover probability of about 30\% is achieved.   As various amounts of cancellation error are considered, $\varepsilon > 0$, there is a nonzero probability that the macro user cannot maintain their required SINR.  In Fig.~\ref{fig:goodlinksMU}, we show on average how many macro users can maintain their required SINR after the handover.  
For cancellation errors $\varepsilon < 12.5$\%, we see that about 98\% of macro users are served by either the BS or FAP.  Recall that we are considering $M = 25$ macro users, thus at most one or two of the users see worse performance after the handover.  

\begin{figure}
  \includegraphics[width=0.48\textwidth]{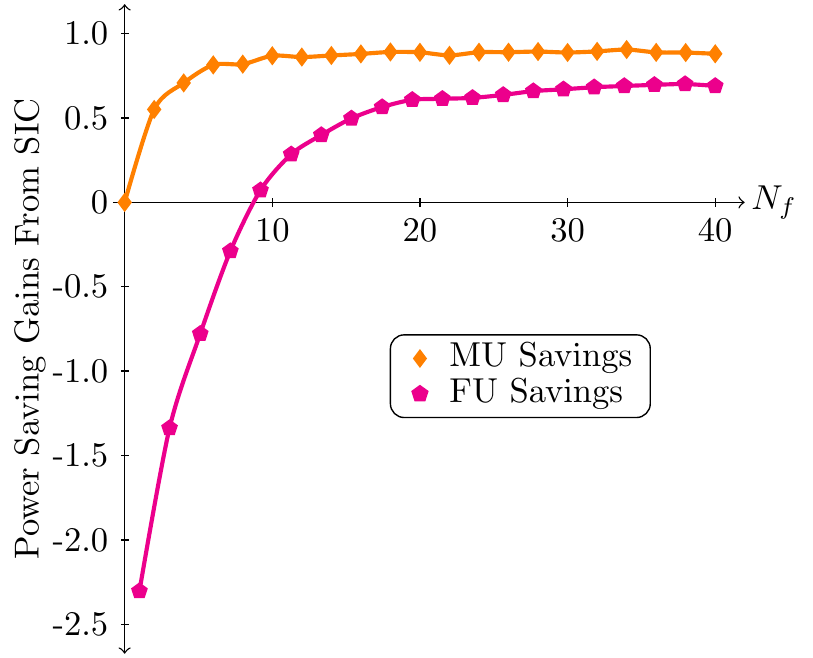}
\caption{Significant power savings are achieved by both macro and femto users by allowing macro users to join nearby femtocells.}
\label{fig:powersavings}
\end{figure}

Because there is some probability that successive interference cancellation will not work perfectly, this shared spectrum handover scheme will always be opportunistic.  However, at the beginning of this section we discussed power savings as a potential benefit for macro users to join a nearby femtocell.  In Fig.~\ref{fig:powersavings},
we plot the power savings for both macro and femto users for the case of $\varepsilon = 0$.  We can immediately see that both users can achieve significant power savings especially for $N_{f} > 15$.  Macro users have the most opportunity to save power as they can connect to a nearby femtocell instead of the potentially distant base station.  Furthermore, macro users will always have a positive gain as they will never join a femtocell if they have to use more power than required to reach the base station.  However, femto users do observe a negative gain when the femtocell density is low, i.e. $N_{f} < 9$.  Femto users have to transmit at a higher level of power in order to share their channel with a macro user.  As the femtocell density increases however, the benefits of the lower macro user power are realized.   

\begin{figure}
  \includegraphics[width=0.48\textwidth]{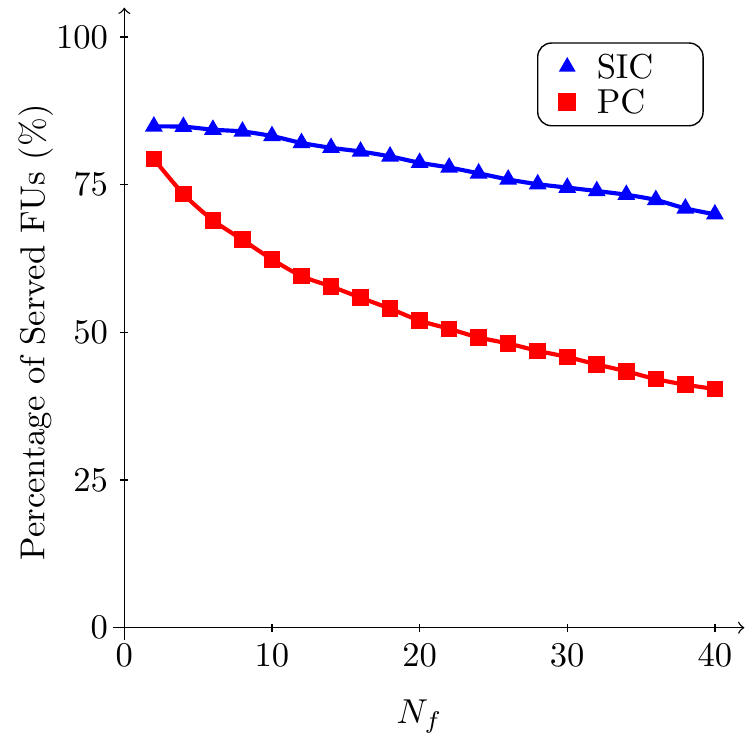}
\caption{Using successive interference cancellation (SIC) to allow macro users to join nearby femtocells outperforms power control alone (PC) in terms of the average number of served femto users. }
\label{fig:goodlinks}
\end{figure}

In addition to power savings, femto user link quality also improves by allowing macro users to join nearby femtocells.  The lower transmit power used by the macro users reduces the total interference at each FAP.  This in turn allows for more femto uses to satisfy their required SINR at the FAP.  In Fig.~\ref{fig:goodlinks}
we plot the average number of femto users who can satisfy the required SINR threshold at their respective FAPs.  We show curves for both the PC scheme without the handover process and the SIC scheme that allows the handover to occur.  We can clearly see that the SIC scheme outperforms the PC scheme and at high values of $N_f$, large gains in the number of users served are realized.  We also note that at smaller values of $N_f$, the PC scheme's performance decays at a faster rate than SIC.  Then around $N_f = 25$, the two schemes begin to decay at the same rate.

As more femto users are served in the network, additional gains in terms of network throughput will be realized.  Recall that the $M$ macro users are always guaranteed a channel of at least an SINR level of $\beta_M$ from either the BS or a nearby FAP.  Thus the macro user component of the sum rate will always be equal to $M \log_2(1+\beta_M)$ whether or not the femtocell network is present.  Any gains in the sum rate will come from the additional femto users that are active in the network.  Based on this, we can write the gain in the sum rate for the shared spectrum scheme as 
\begin{equation}
\mathcal{R}_{gain}^{share} = \dfrac{\mathcal{R}_{F}}{M\log_2(1+\beta_M)},
\end{equation}
where the gains are calculated as a percentage of the sum rate of the macro user only network.  We note that $\mathcal{R}_{F}$ is calculated in the same way as in (\ref{eq:frate}).  We can derive an upper bound on the sum rate gain from the scenario that all femto users in each femtocell are able to satisfy their required SINR threshold with their corresponding femtocell access point.  We know that there are on average $FN_f$ femto users per macrocell thus it is easy to show that the maximum sum rate gain satisfies the condition
\begin{equation}
\mathcal{R}_{gain}^{share} < \dfrac{FN_f \log_2(1+\beta_F)}{M\log_2(1+\beta_M)} \triangleq \mathcal{R}_{max},
\label{eq:upperrate}
\end{equation}
where the upper bound is linear in the average number of femtocells per macrocell.  
\begin{figure}
  \includegraphics[width=0.48\textwidth]{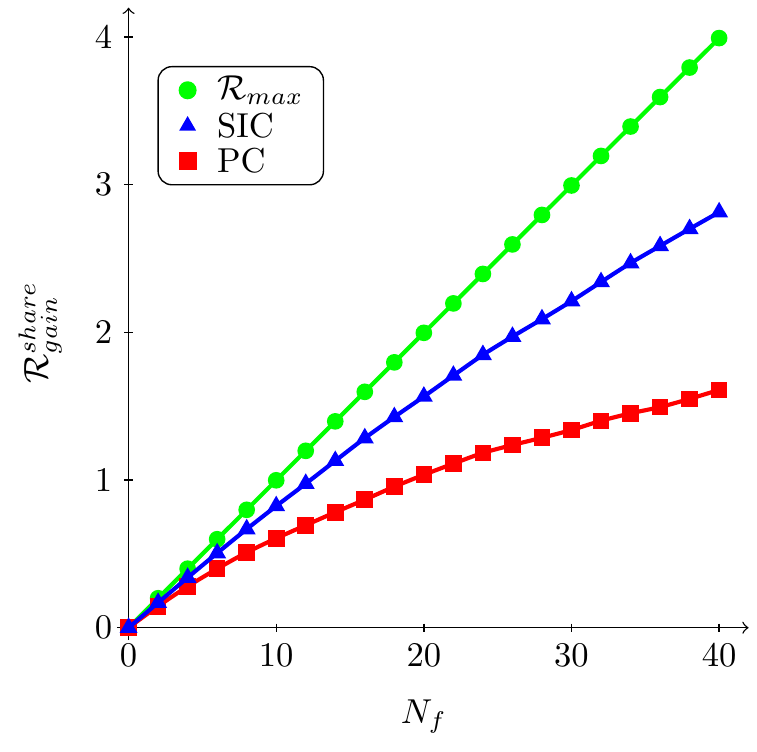}
\caption{The average network sum rate gain ($\mathcal{R}_{gain}^{share}$) versus the average number of femtocells ($N_{f}$). Both the power control only (PC) and successive interference cancellation (SIC) schemes are presented. }
\label{fig:tputgainshare}
\end{figure}

In Fig.~\ref{fig:tputgainshare} we plot the average sum rate gain of the network for the two schemes considered above.  In addition, we plot the upper bound on the sum rate gain as found in (\ref{eq:upperrate}).  We can immediately see that the SIC scheme outperforms the PC scheme in terms of the sum rate gain achieved.  As $N_f$ increases, we see that the amount of gain of the SIC scheme over PC scheme also increases.  We further note that the SIC scheme is significantly closer to the maximum sum rate gain than the PC scheme for all values of $N_F > 10$.

\section{Conclusions}
\label{sec:conclusions}
In this work, we have presented two architectures for an underlaid femtocell network and analyzed the performance from the perspective of both the whole network and individual users.  The first scheme allocates orthogonal partitions of the network bandwidth to the macrocell and femtocell networks.  In doing so, the interference management becomes much simpler as the femto users can utilize the spectrum without interfering with the cellular base station or receiving interference from potentially high powered macro users.  Inter-femtocell interference is the only limiting factor in both network and user performance.  We measure the average femto user performance by the expected SINR and show that for pathloss exponent $\alpha \geq 2$, femto users can maintain high link qualities and performance decreases linearly with the femtocell density.  The performance of the network as a whole also increases with sum rate gains up to 200\%.  

We then presented a shared spectrum femtocell architecture in the second part of our work.  Interference management is much more difficult as femto users have to limit the interference they cause to the cellular base station and avoid the interference they receive from nearby macro users.  We developed a femto user power control scheme that relies on minimal coordination with the cellular base station allowing femtocells to exist and operate independently of each other.  Femtocell access points then use an interference sensing scheme to allocate channels to its users.  These two techniques alone can give sum rate gains up to 200\%.  

We then developed a decision rule for macro users to decide whether to connect to their own cellular base station or a nearby femtocell access point.  Successive interference cancellation is used to allow a macro and femto user to share a single channel and connect to the femtocell access point in a multiple access manner.  Our results showed that about 30\% of macro users satisfy the condition to join a femtocell.  In doing so, macro users can save up to 90\% of their power instead of connecting to the potentially distant base station.  As a result of sharing their channel however, femto users may have to actually use more power to overcome the macro user's interference.  For very sparse femtocell networks, nine or less femtocells, there are no power savings and in fact a large increase in power consumption.  However, once the femtocell density increases, femto user power savings quickly approaches 70\%.  Allowing macro users to join nearby femtocells increases the performance of the network as a whole.  Specifically, sum rate gains up to 300\% can be achieved due to the overall lower interference level in the network.  
\ifCLASSOPTIONcaptionsoff
  \newpage
\fi



%
%
%

\bibliographystyle{IEEEtran} 
\bibliography{femto.bib}

\begin{thebibliography}{10}
\providecommand{\url}[1]{#1}
\def\UrlFont{\rmfamily}
\providecommand{\newblock}{\relax}
\providecommand{\bibinfo}[2]{#2}
\providecommand\BIBentrySTDinterwordspacing{\spaceskip=0pt\relax}
\providecommand\BIBentryALTinterwordstretchfactor{4}
\providecommand\BIBentryALTinterwordspacing{\spaceskip=\fontdimen2\font plus
\BIBentryALTinterwordstretchfactor\fontdimen3\font minus
  \fontdimen4\font\relax}
\providecommand\BIBforeignlanguage[2]{{%
\expandafter\ifx\csname l@#1\endcsname\relax
\typeout{** WARNING: IEEEtran.bst: No hyphenation pattern has been}%
\typeout{** loaded for the language `#1'. Using the pattern for}%
\typeout{** the default language instead.}%
\else
\language=\csname l@#1\endcsname
\fi
#2}}

\bibitem{FCC_AUCTION}
FCC, ``Auction of 700 {MH}z band licenses,'' Press Release, Aug. 2007.

\bibitem{Claussen2008An-Overview-of-}
H.~Claussen, L.~T.~W. Ho, and L.~G. Samuel, ``An overview of the femtocell
  concept,'' \emph{Bell Labs Tech. J.}, vol.~13, no.~1, pp. 221--246, 2008.

\bibitem{Femtocell_Survey}
V.~Chandrasekhar, J.~G. Andrews, and A.~Gatherer, ``Femtocell networks: A
  survey,'' \emph{IEEE Comm. Mag.}, vol.~46, no.~9, pp. 59--67, Sep. 2009.

\bibitem{3GPPstan}
3GPP, ``{UTRAN} architecture for 3{G} home {N}ode {B} ({HNB}); {S}tage 2,'' TS
  25.467 (release 9), 2010.

\bibitem{femtoforum}
FemtoForum, ``Interference management in {OFDMA} femtocells,'' available at
  www.femtoforum.org, Tech. Rep., Mar 2010.

\bibitem{lopez_interference}
D.~Lopez-Perez, A.~Valcarce, G.~de~la Roche, and J.~Zhang, ``{OFDMA}
  femtocells: {A} roadmap on interference avoidance,'' \emph{IEEE Comm. Mag.},
  vol.~47, no.~9, pp. 41--48, Sep. 2009.

\bibitem{andrews_int_can}
J.~G. Andrews, ``Interference cancellation for cellular systems: A contemporary
  overview,'' \emph{IEEE Wireless Comm. Mag.}, vol.~12, no.~2, pp. 19--29, Apr.
  2005.

\bibitem{andrews_powercontrol}
V.~Chandrasekhar, J.~G. Andrews, T.~Muharemovic, Z.~Shen, and A.~Gatherer,
  ``Power control in two-tier femtocell networks,'' \emph{IEEE Wireless Comm.},
  vol.~8, no.~8, pp. 4316--4328, Aug. 2009.

\bibitem{Han-Shin-Jo2009Interference-Mi}
H.-S. Jo, C.~Mun, J.~Moon, and J.-G. Yook, ``Interference mitigation using
  uplink power control for two-tier femtocell netwroks,'' \emph{IEEE Trans.
  Wireless Commun.}, vol.~8, no.~10, pp. 4906--4910, Oct. 2009.

\bibitem{hybrid_freqassign}
I.~Guvenc, M.-R. Jeong, F.~Watanabe, and H.~Inamura, ``A hybrid frequency
  assignment for femtocells and coverage area analysis for co-channel
  operation,'' \emph{IEEE Comm. Letters}, vol.~12, no.~12, Dec. 2008.

\bibitem{FFR_infocom}
A.~Stolyar and H.~Viswanathan, ``Self-orgainizing dynamic fractional frequency
  reuse in {OFDMA} systems,'' in \emph{Proc. IEEE INFOCOM}, Apr. 2009.

\bibitem{int_can_ahls}
R.~Ahlswede, ``Multi-way communication channels,'' in \emph{Proc. IEEE Int.
  Symp. Inform. Th.}, Sep. 1971, pp. 23--52.

\bibitem{int_can_4G}
G.~Boudreau, J.~Panicker, N.~Guo, R.~Chang, N.~Wang, and S.~Vrzic,
  ``Interference coordination and cancellation for 4{G} networks,'' \emph{IEEE
  Comm. Mag.}, vol.~47, no.~4, pp. 74--81, Apr. 2009.

\bibitem{SIC_OFDM}
Y.~Whang, J.-H. Park, and R.~Whang, ``Low complexity successive interference
  cancellation for {OFDM} systems over time-varying multipath channels,'' in
  \emph{Proc. IEEE Vehicular Techonlogy Conf.}, Jun. 2009.

\bibitem{Martin-Sacristan2009On-the-Way-Towa}
D.~Martin-Sacristan, J.~Monserrat, J.~Cabrejas-Penuelas, D.~Calabuig,
  S.~Garrigas, and N.~Cardona, ``On the way towards fourth-generation mobile:
  {3GPP LTE} and {LTE}-advanced,'' \emph{EURASIP Journal on Wireless Commun.
  and Networking}, 2009.

\bibitem{bkaufman_j1}
B.~Kaufman, J.~Lilleberg, and B.~Aazhang, ``Spectrum sharing between cellular
  networks and ad-hoc device to device networks,'' \emph{under review in IEEE
  Trans. on Wireless Commun.}, May. 2012.

\bibitem{cover_thomas}
T.~M. Cover and J.~A. Thomas, \emph{Elements of Information Theory}.\hskip 1em
  plus 0.5em minus 0.4em\relax John Wiley {\&} Sons, 1991.

\end{thebibliography}

\end{document}